\documentclass[11pt,tightenlines,preprint,amssymb,showpacs,nofootinbib,eqsecnum,floatfix,byrevtex]{revtex4}
\usepackage{eucal}
\usepackage{color}
\usepackage{graphics}
\usepackage{amsmath}

\def\be{\begin{equation}}
\def\ee{\end{equation}}
\def\bea{\begin{eqnarray}}
\def\eea{\end{eqnarray}}

\newcommand{\sq}[1]{\sigma^{(q)}_{[D_{#1}]}}

\newcommand{\sd}[1]{\sigma^{(d)}_{[D_{#1}]}}
\newcommand{\sD}[1]{\hat{\sigma}^{(d)}_{[D_{#1}]}}
\newcommand{\sS}{\sigma_{[A1]}}
\newcommand{\cu}[1]{c^{(u)}_{#1}}
\newcommand{\cd}[1]{c^{(d)}_{#1}}
\newcommand{\de}[2]{(\delta_{#1}^d)_{#2}}
\newcommand{\deu}[2]{(\delta_{#1}^u)_{#2}}
\newcommand{\cpu}[1]{c^{\prime(u)}_{#1}}
\newcommand{\cpd}[1]{c^{\prime(d)}_{#1}}

\begin{document}
\preprint{\sf KUNS-2066, OHSTPY-HEP-T-06-001, P07015, TU-786}

\title{String-derived $D_4$ flavor symmetry and phenomenological implications}

\author{{\sc Pyungwon Ko}}
\affiliation{School of Physics, KIAS, Cheongnyangni-dong,
Seoul, 130--722, Korea }
\author{{\sc Tatsuo Kobayashi}}
\affiliation{Department of Physics, Kyoto University, Kyoto 606-8502, Japan.}
\author{{\sc Jae-hyeon Park}}
\affiliation{Department of Physics, Tohoku University, Sendai 980--8578,
Japan.}
\author{{\sc Stuart Raby}}
\affiliation{Department of Physics, The Ohio State University,
Columbus, OH 43210 USA.}

\begin{abstract}
In this paper we show how some flavor symmetries may be derived
from the heterotic string, when compactified on a  6D orbifold.
In the body of the paper we focus on the $D_4$ family symmetry,
recently obtained in $Z_3 \times Z_2$ orbifold constructions.
We show how this flavor symmetry constrains fermion masses,
as well as the soft SUSY breaking mass terms.  Flavor symmetry breaking
can generate the hierarchy of fermion masses and at the same time the
flavor symmetry suppresses large flavor changing neutral current processes.
%
\end{abstract}
\date{April 20, 2007}

\maketitle

\section{Introduction}

Fermion masses and mixing angles are the precision low energy data
which will test any new physics beyond the standard model.
Quarks and leptons come in three flavors (or families) with a distinct
hierarchy of masses for charged fermions; with the third family heavier
than the second, which is heavier than the first.   Moreover the
mixing angles evident in charge current electroweak processes are small
and favor nearest neighbor mixing with respect to family number.
In the neutrino sector, the situation is not as clear.
There are possibly more than three light neutrinos.   Their masses may be
Majorana or Dirac.  Even if one assumes just three light Majorana
neutrinos, a normal or inverted hierarchy is possible.  Finally
the leptonic mixing angles, so-called PMNS angles, (analogs of the
CKM mixing for quarks) are large, with maximal mixing ($\sim 45^\circ$)
between $\nu_\mu - \nu_\tau$ and large mixing ($\sim 30^\circ$) between
$\nu_e - (\nu_\mu, \nu_\tau)$.  Within the context of the See-Saw mechanism,
this may be generated via large mixing in the Dirac neutrino mass matrix
or in the right-handed Majorana mass matrix~\cite{lopsided}.
However Ref.~\cite{Dermisek:2004tx} showed that bi-large neutrino
mixing may also be obtained with completely hierarchical Dirac and Majorana
neutrino mass matrices.  In a recent paper this analysis was extended to
supersymmetric SO(10)~\cite{Dermisek:2005ij}.

Understanding the origin of fermion flavor structure, i.e. fermion
masses and mixing angles, is one of the important issues in particle
physics. In field-theoretical model building, one often assumes
certain types of flavor symmetries, which control Yukawa couplings
and higher dimensional operators. Higher dimensional operators are
useful as effective Yukawa couplings when proper scalar fields
develop their expectation values (VEVs). Continuous and discrete
non-abelian symmetries, e.g. $U(2)$, $D_4$, $S_3(\approx D_3)$,
$A_4$, $Q_n$, $\Delta(3n^2)$, $\Delta(6n^2)$ are assumed as flavor
symmetries
\cite{Dermisek:2005ij,Frampton:1994rk,flavor-u2,flavor-s3,flavor-A4,
flavor-d4,Babu:2004tn,Hagedorn:2006ir,Kaplan:1993ej}, while abelian
symmetries such as $U(1)$, $Z_N$ are often assumed, too. However,
from the viewpoint of 4D field theory their origins are not clear.

Such flavor structure also has significant implications for supersymmetry
(SUSY) breaking terms\cite{Dine:1993np}. Superpartners have not been
detected yet, but flavor changing neutral current (FCNC)
processes are strongly constrained by experiments
\cite{Gabbiani:1996hi,fcnc};\footnote{See also \cite{Chankowski:2005jh}
and references therein.} requiring that soft SUSY breaking terms should be
approximately degenerate between the first and second families.
Thus, in several field-theoretical models the first and second families
are assumed to be a doublet under certain flavor symmetries,
while the third family may be a singlet.
Furthermore, there are FCNC constraints for the third family,
although not as restrictive as for the first and second families.

Superstring theory is a promising candidate for a unified theory
including gravity. Hence, it is very important to study what type of flavor
structures can be realized within the framework of string models and
to investigate their implications for particle physics.
Heterotic orbifold models can lead to realistic 4D models, and one
interesting feature is that phenomenological aspects are determined
by geometrical properties of the orbifolds. For examples,
in Ref.~\cite{Ibanez:1987sn,Font:1988mm,Casas:1988hb,Casas:1988vk}
$Z_3$ models with three families have been obtained.
In $Z_3$ orbifold models, the untwisted matter sector has the degeneracy
factor three, which originates from a triplet of $SU(3)_H$ holonomy;
a sub-group of broken $SU(4)_R$ symmetry. Also each 2D $Z_3$ orbifold
of a 6D orbifold has three fixed points, with a degenerate massless spectra,
unless one introduces Wilson lines. These triplets in the untwisted and
twisted sectors correspond to three families in the models of
Ref.~\cite{Ibanez:1987sn,Font:1988mm,Casas:1988hb,Casas:1988vk}.
In this case, the Higgs fields must also be a triplet such that at least
the top Yukawa coupling is allowed as a 3-point coupling.

Recently, a new type of model has been constructed on the
$Z_2 \times Z_3 = Z_6$ orbifold \cite{Kobayashi:2004ud,Kobayashi:2004ya}.
\footnote{ See also for recent studies on model construction
\cite{Forste:2004ie}.}
In these models, three families are realized as a singlet and doublet
under a $D_4$ flavor symmetry. That is these models are, as far as we know,
the first models to realize such flavor structure. The third family of
quarks and leptons are $D_4$-singlets. Thus, one important aspect of such
flavor structure is that the Higgs field, which is allowed to couple to
the top quark, is a singlet under the flavor symmetry.
Moreover, the flavor structure has other significant implications in the
Yukawa matrices and SUSY breaking terms. Hence, it is important to study
phenomenological aspects of the string-derived $D_4$ flavor
symmetry as well as other discrete non-abelian flavor symmetries.

Furthermore, in Ref.~\cite{Kobayashi:2006wq}, all possible
non-Abelian discrete flavor symmetries, which can appear from
heterotic orbifold models, have been classified. Those include
$D_4$, $\Delta(54)$ and $SW_4$, and $\Delta(54)$ can break into
$D_3$. In addition, it has been shown that the model on $S^1/Z_2$
can include only $D_4$-doublets and trivial singlets as fundamental
modes, but models on $T^2/Z_4$ and $T^4/Z_8$ can include non-trivial
$D_4$ singelts, too.

Here we study the $D_4$ flavor structure.  Generic heterotic orbifold
models leading to the $D_4$ flavor structure are considered.
We also analyze their phenomenological aspects, that is, effective Yukawa
couplings and SUSY breaking terms.
Furthermore we study FCNC constraints on SUSY breaking terms in detail.

This paper is organized as follows. In section II, we show how the
$D_4$ flavor structure appears from heterotic orbifold models. In
section III, we study its implication on Yukawa matrices. In section
IV, we study predictions of SUSY breaking squark masses and scalar
trilinear couplings. Section V is devoted to conclusion and
discussion. In Appendix A, group-theoretical aspects of $D_4$ are
summarized. In Appendix B, we summarize sfermion masses and scalar
trilinear couplings, derived from a $U(1)$ Froggatt-Nielsen model to
compare with our model. In Appendix C, we comment on a possibility
to realize SUSY breaking terms consistent with
$t-b-\tau$ Yukawa unification.

\section{$D_4$ flavor structure in string models}

\subsection{$D_4$ flavor symmetry from orbifold models}

In this section, we show that the $D_4$ flavor structure can be derived
in $Z_2 \times Z_M$ heterotic orbifold models.
(See also \cite{Kobayashi:2006wq}.)
The $Z_2 \times Z_M$ orbifold is obtained as follows. First, we consider
the 6D $T^2 \times T^2 \times T^2$ torus. Then, we divide it by two
independent twists, $\theta$ and $\omega$, whose eigenvalues are $e^{2\pi i
v_1}$ and $e^{2\pi i v_2}$ with
\begin{equation}
v_1 = (1/2,-1/2,0), \qquad v_2 = (0,1/M,-1/M),
\end{equation}
respectively, in the complex basis, such that we obtain the
$Z_2 \times Z_M$ orbifold preserving N=1 4D SUSY. Note that the first
plane becomes the 2D $Z_2$ orbifold by dividing the $T^2$ by the $Z_2$
twist. The 2D $Z_2$ orbifold has four fixed points denoted as
$(\frac{n_1}{2}{\bf e}_1+\frac{n_2}{2}{\bf e}_2)$ for $n_1,n_2=0,1$,
where ${\bf e}_1$ and ${\bf e}_2$ are lattice vectors defining $T^2$.
The twisted states are associated with these fixed points.
Namely, the $\theta^{2k+1} \omega^\ell$ twisted states have the degenerate
spectrum for these four fixed points, while for  the
$\theta^{2k} \omega^\ell$ twist this space is just the fixed torus and the
degeneracy factor from this space is just one.

We can introduce degree-two Wilson lines $2W_i = \Gamma_G$
associated with ${\bf e}_i$ ($i=1,2$), where $\Gamma_G$ is the gauge
lattice, i.e. the $E_8 \times E_8$ lattice for the $E_8 \times E_8$
heterotic string theory. For example, the non-vanishing Wilson line
$W_1$ resolves the degeneracy between the fixed points
$\frac{n_2}{2}{\bf e}_2$ and $\frac{1}{2}{\bf e}_1
+\frac{n_2}{2}{\bf e}_2$ ($n_2=0,1$), that is, the massless spectrum
of the twisted states corresponding to the fixed points
$\frac{n_2}{2}{\bf e}_2$ differs from one corresponding to the fixed
points $\frac{1}{2}{\bf e}_1 +\frac{n_2}{2}{\bf e}_2$ ($n_2=0,1$).
However, there still remains the degeneracy factor two unless we
introduce a non-trivial Wilson line $W_2$ along the ${\bf e}_2$
direction. Thus, the degeneracy between the states $| n_2 =0 \rangle
$ and $| n_2 =1 \rangle $ in the $\theta^{2k+1} \omega^\ell$ twisted
sector is the origin of doublets under our flavor symmetry, studied
in this paper. We will show later that the flavor symmetry is
actually the $D_4$ symmetry. On the other hand, the $\theta^{2k}
\omega^\ell$ twisted sector as well as the untwisted sector does not
have such a fixed point structure. Hence, the states in such sectors
correspond to a singlet under our flavor symmetry.

Here we show that the above flavor structure corresponds to the
$D_4$ flavor symmetry. The Lagrangian has the permutation symmetry
between the states $|n_2 \rangle$ with $n_2=0,1$. In addition, each
coupling is controlled by the $Z_2$ symmetry, under which the state
$|n_2 \rangle$ is transformed as $|n_2 \rangle \rightarrow
(-1)^{n_2} |n_2 \rangle$. These transformations are denoted by the
two Pauli matrices,
\begin{equation}
\sigma_1 = \left(
\begin{array}{cc}
0 & 1 \\ 1 & 0 \\
\end{array}
\right), \qquad
\sigma_3 = \left(
\begin{array}{cc}
1 & 0 \\ 0 & -1 \\
\end{array}
\right),
\end{equation}
respectively on the state basis $(| n_2 =0 \rangle, | n_2 =1 \rangle )$.
The complete closed set of operations
forms the discrete non-abelian $D_4$ symmetry, which consists of
\begin{equation}
\pm I, \qquad \pm \sigma_1, \qquad \pm i\sigma_2,
\qquad \pm \sigma_3 .
\end{equation}
The $D_4$ symmetry is a symmetry of a square.
Thus, the $\theta^{2k+1}\omega$ twisted states are $D_4$ doublets,
while the other $\theta^{2k}\omega$ twisted states and the untwisted
states are $D_4$ singlets.

Note, a field theoretical orbifold GUT explanation may be useful for
field theory model builders. We consider the model with the extra
dimension $S^1/Z_2$, which has two fixed points. String theory
requires that brane fields on these two fixed points must be
degenerate in the massless spectrum unless a non-vanishing Wilson
line is introduced to resolve this degeneracy. Brane fields on two
fixed points of $S^1/Z_2$ are $D_4$ doublets corresponding to
$\theta^{2k+1}\omega$ twisted states in $Z_2 \times Z_M$ heterotic
orbifold models. On the other hand, bulk fields on $S^1/Z_2$ are
$D_4$ singlets corresponding to $\theta^{2k}\omega$ twisted states
and untwisted states.
Finally string selection rules
require that the superpotential contain an even number of doublet
fields at each fixed point.

\subsection{Explicit model \label{sec:explicitmodel}}

Here we give the example with the $D_4$ flavor structure, which has been
obtained in Ref.~\cite{Kobayashi:2004ya}. That is the $Z_2\times Z_3 = Z_6$
orbifold model with the Pati-Salam gauge group
$SU(4) \times SU(2)_L \times SU(2)_R$ and the extra gauge group
$SO(10)'\times SU(2)' \times U(1)^5$. The model has three families
under the Pati-Salam group, i.e.
$3 \times [({\bf 4},{\bf 2},{\bf 1})+ (\bar {\bf 4},{\bf 1},{\bf 2})]$.
The untwisted sector has one family, $({\bf 4},{\bf 2},{\bf 1})+
(\bar {\bf 4},{\bf 1},{\bf 2})$ under $SU(4) \times SU(2)_L \times SU(2)_R$,
which include the third family of left-handed quarks and anti-quarks,
$q_3, \bar u_3$ and $\bar d_3$. The $\theta \omega$ twisted sector has
the other two families, which include the first and second families of
quarks, $q_i, \bar u_i$ and $\bar d_i$ ($i=1,2$).
The higgs field $({\bf 1},{\bf 2},{\bf 2})$ comes from the untwisted sector,
and it includes the up-sector and down-sector higgs fields,
$h_u$ and $h_d$. Thus, the third family of quarks as well as higgs fields
are $D_4$ singlets, while the other two families are $D_4$ doublets.
In Table 1, we show their extra $U(1)$ charges for later convenience.
This model also includes extra matter fields usually found in string models.
(See for details Ref.~\cite{Kobayashi:2004ya}.)

\begin{table}[htb]
\begin{center}
\begin{tabular}{|c|c|c|c|c|c|}
\hline
& $Q_1$ & $Q_2$ & $Q_3$ & $Q_4$ & $Q_A$ \\
\hline \hline $q_3$ & 1 & 1 & 0 & 3 & --2 \\
\hline $q_{1,2}$ & --1 & 0 & 0 & 0 & 0 \\
\hline $\bar u_3, \bar d_3$ & --3 & 0 & 0 & --1 & 0 \\
\hline $\bar u_{1,2}, \bar d_{1,2}$ & --1 & 0 & 0 & 0 & 0 \\
\hline $h_u, h_d$ & 2 & --1 & 0 & --2 & 2 \\
\hline \end{tabular} \end{center}
\caption{Extra U(1) charges in explicit string model} \end{table}

The discussion in the rest of the paper is quite generic and
independent of the particular gauge group, since we are now more
interested in the consequences of the flavor symmetries.  Therefore,
we will discuss the case that three families of quarks and leptons
in the basis of the Standard Model gauge group consist
of singlets and doublets under the $D_4$ flavor symmetry. In the
following sections, we concentrate on the phenomenological
implications of the $D_4$ flavor structure in the quark sector.

\section{Yukawa matrices}

In this section and the next section, we study phenomenological
implications of the $D_4$ flavor structure. First, in this section
we consider Yukawa matrices. We consider the $D_4$ flavor structure
where the third family $q_3,\bar u_3,\bar d_3$
corresponds to the $D_4$-trivial singlet $A_1$ and the first and
second families $q_i,\bar u_i,\bar d_i$ $(i=,1,2)$ are
$D_4$-doublets. The up- and down-sectors of higgs fields are also
$D_4$-singlets.  (See Appendix A for more details on the discrete
group $D_4$ and its representations.)

Let us examine the 3-point couplings,
\begin{equation}
y^{(u)}_{ij} q_i \bar u_j H_u, \qquad y^{(d)}_{ij} q_i \bar d_j H_d .
\end{equation}
The $D_4$ algebra allows diagonal entries, i.e.
$y^{(u,d)}_{ij} = y^{(u,d)}_i \delta_{ij}$ with $y^{(u,d)}_1 =
y^{(u,d)}_2 \neq y^{(u,d)}_3$. If $y^{(u,d)}_i =O(1)$, that is not realistic.
Thus, we assume that extra symmetries allow the couplings of
the third family, but not the first or second families, i.e.
\begin{equation}
y^{(u,d)} = \left(
\begin{array}{ccc}
 0& 0& 0 \\
 0& 0& 0 \\
 0& 0& y^{(u,d)}_{33}
\end{array}
\right).
\label{yukawa-1}
\end{equation}
Actually, the example shown in Section \ref{sec:explicitmodel} has
extra $U(1)$ symmetries which forbid the Yukawa couplings for the
first and second families, but allow the third family Yukawa
couplings. In the explicit string model,  other stringy selection
rules also forbids the Yukawa couplings for the first and second
families \cite{Kobayashi:2004ya}. Furthermore, in the model
\cite{Kobayashi:2004ya}, the third family Yukawa couplings are
required to be the same as the gauge coupling $g$, i.e.
$y^{(u,d)}_{33} = g \approx O(1)$.

Now, let us consider how to generate the other entries
of the Yukawa matrix. Those are expected to be
generated as effective Yukawa couplings through higher dimensional
operators once certain scalar fields develop their VEVs. String
models, in general, have several gauge-singlet fields $\sigma$, and
they can develop VEVs along flat directions. Of course, each
gauge-singlet field transforms as a trivial singlet $A_1$ or a
doublet under the $D_4$ group. In addition, a product of
$D_4$-doublets include four types of $D_4$-singlets, $A_1$, $B_1$,
$B_2$ and $A_2$. Thus, higher dimensional operators  can generate
the following effective Yukawa matrices,
\begin{equation} y^{(u,d)} = \left( \begin{array}{ccc}  \sigma^{(u,d)}_{[A_1]} + \sigma^{(u,d)}_{[B_2]} &
 \sigma^{(u,d)}_{[B_1]} + \sigma^{(u,d)}_{[A_2]} &
\sigma^{(u,d)}_{[D_1]} \\
 \sigma^{(u,d)}_{[B_1]} - \sigma^{(u,d)}_{[A_2]} &
 \sigma^{(u,d)}_{[A_1]} - \sigma^{(u,d)}_{[B_2]} &
\sigma^{(u,d)}_{[D_2]} \\
 \sigma^{(u,d)}_{[D'_1]} & \sigma^{(u,d)}_{[D'_2]} & 1 \end{array} \right),
\label{yukawa-2} \end{equation} up to $O(1)$ coefficients (assuming
the singlet fields have proper $U(1)$ charges). Here,
$\sigma^{(u,d)}_{[R]}$ denotes a product of gauge singlets in the
$R$ representation under $D_4$, and $\sigma^{(u,d)}_{[D_1]}$ and
$\sigma^{(u,d)}_{[D_2]}$ are $D_4$-doublet.\footnote{In the explicit
model \cite{Kobayashi:2004ya} we can generate such effective Yukawa
matrices by SM gauge singlets, contained in the model.} Here, their
VEVs are denoted by dimensionless parameters with units $M =1$,
where $M$ is the Planck scale. These effective Yukawa matrices have
more than enough free parameters (as VEVs of singlets $\sigma$),
such that one can realize realistic values of quark masses and
mixing angles in a generic model. In other words, we have no
prediction in the Yukawa sector using only the $D_4$ flavor
symmetry, unless additional symmetries or conditions for $\sigma$
are imposed.   Note, for later reference, the
experimental values of the quark mass ratios and mixing angles can
be given by the approximate relations,
\begin{eqnarray}
 & & \frac{m_d}{m_s} \sim \lambda^2, \qquad
\frac{m_s}{m_b} \sim \frac{1}{2} \lambda^2, \\
 & & \frac{m_u}{m_c} \sim \lambda^4, \qquad
\frac{m_c}{m_t} \sim  \lambda^3 - \lambda^4, \\
 & & V_{us} \sim \lambda, \qquad V_{cb} \sim \lambda^2,
\qquad V_{ub} \sim  \lambda^3 - \lambda^4,
\end{eqnarray}
where $\lambda = 0.22$.

\vskip 1cm {\bf Model}

\vskip 0.3cm Here we show a simple example leading to realistic
results. We introduce $\sq{1,2}$, $\sd{1,2}$, $\sD{1,2}$ and $\sS$
(unrelated to the variables $\sigma^{(u,d)}$ introduced in Eqn.
\ref{yukawa-2}), and assume they develop vevs.  $\sS$ is a $D_4$
trivial singlet and the others are $D_4$ doublets.  We also
introduce two extra $U(1)$ symmetries.  Note, the $U(1)$ charges of
each field are assigned in Table \ref{tab:u1}. The following
discussion is independent of values of the $U(1)$ symmetry
parameters $a$,$b$, $x$ and $y$ (see Table \ref{tab:u1}). A working
example is
\begin{equation}
  a = 1, \quad b = 0, \quad x = 0, \quad y = 1 .
\end{equation}
Then we obtain the following forms of Yukawa matrices,
\begin{equation}
  \begin{aligned}
  y^{(u)} &= \left(
  \begin{array}{cc|c}
    \cu{a} \sd{1} \sD{1} + \cu{b} \sd{2} \sD{2} &
    \cu{b} ( \sd{1} \sD{2} + \sd{2} \sD{1} )    & \cu{3} \sq{1} \\
    \cu{b} ( \sd{1} \sD{2} + \sd{2} \sD{1} )    &
    \cu{b} \sd{1} \sD{1} + \cu{a} \sd{2} \sD{2} & \cu{3} \sq{2} \\
    \hline
    0       & 0       & 1
  \end{array}
  \right) , \\
  y^{(d)} &= \left(
  \begin{array}{cc|c}
    \cd{a} \sd{1} \sD{1} + \cd{b} \sd{2} \sD{2} &
    \cd{b} ( \sd{1} \sD{2} + \sd{2} \sD{1} )    & \cd{3} \sq{1} \sS \\
    \cd{b} ( \sd{1} \sD{2} + \sd{2} \sD{1} )    &
    \cd{b} \sd{1} \sD{1} + \cd{a} \sd{2} \sD{2} & \cd{3} \sq{2} \sS \\
    \hline
    0       & 0       & \sS
  \end{array}
  \right) .
  \end{aligned}
\end{equation}
Here the (3,1) and (3,2) entries in both Yukawa
matrices of up and down sectors are quite suppressed, and
irrelevant to the following discussions.

\begin{table}
  \centering
  \begin{tabular}{c|c|c}
    \hline
    & $Q_1$ & $Q_2$ \\
    \hline
    $q_3$           & 1           & 3 \\
    $q_{1,2}$       & $-1$        & 0 \\
    $\bar{u}_3$     & $-3$        & $-1$ \\
    $\bar{u}_{1,2}$ & $-1$        & $0$ \\
    $\bar{d}_3$     & $-3 - a$    & $-1 - b$ \\
    $\bar{d}_{1,2}$ & $-1$        & 0 \\
    $h_u$           & 2           & $-2$ \\
    $h_d$           & 2           & $-2$ \\
    $\sq{1,2}$      & 2           & 3 \\
    $\sd{1,2}$      & $2 + x$     & $3 + y$ \\
    $\sD{1,2}$      & $-2 - x$    & $-1 - y$ \\
    $\sS$           & $a$         & $b$ \\
    \hline
  \end{tabular}
  \caption{U(1) charges of the fields.}
  \label{tab:u1}
\end{table}

We choose the vevs of the $\sigma$ fields as
\begin{equation}
  \label{eq:vevs}
  \begin{aligned}
  \sq{1} &\sim \lambda^3 , \quad
  \sq{2} \sim \lambda^2 , \\
  \sd{1} &\sim \sd{2} \sim \sD{1} \sim \sD{2} \sim \lambda^2 , \\
  \sS &\sim \lambda .
  \end{aligned}
\end{equation}
Then, we have naturally the following texture,
\begin{equation}
  \begin{aligned}
      y^{(u)} & \sim \left(
  \begin{array}{ccc}
    \lambda^4 & \lambda^4 & \lambda^3 \\
    \lambda^4 & \lambda^4 & \lambda^2 \\
    0       & 0       & 1
  \end{array}
  \right) , \qquad
      y^{(d)} & \sim \left(
  \begin{array}{ccc}
    \lambda^4 & \lambda^4 & \lambda^4 \\
    \lambda^4 & \lambda^4 & \lambda^3 \\
    0       & 0       & \lambda
  \end{array}
  \right) .
  \end{aligned}
\end{equation}
This texture can fit the quark mass ratios, the CKM
mixing angles and the KM phase. Note, however, the mass ratios,
$m_u/m_c$ and $m_d/m_s$, and the mixing angle $V_{us}$, are expected
naturally to satisfy $m_u/m_c \approx m_d/m_s \approx V_{us} =O(1)$
unless we fine-tune coefficients. The upper left $2 \times 2$
submatrices are in democratic forms. To realize the mass hierarchy
between the first and the second family quarks, we need the
following fine-tuning of the upper left $2 \times 2$ submatrices for
the up and down sectors, $y^{(u)}_{(22)}$ and $y^{(d)}_{(22)}$,
\begin{equation}
  \begin{aligned}
      y^{(u)}_{(22)} & =  O(\lambda^4) \left(
  \begin{array}{cc}
    1+\varepsilon^{(u)}_{11} & 1+\varepsilon^{(u)}_{12}  \\
   1+\varepsilon^{(u)}_{12} & 1+\varepsilon^{(u)}_{22}
  \end{array}
  \right) , \qquad
      y^{(d)}_{(22)} & =  O(\lambda^4) \left(
  \begin{array}{cc}
    1+\varepsilon^{(d)}_{11} & 1+\varepsilon^{(d)}_{12}  \\
   1+\varepsilon^{(d)}_{12} & 1+\varepsilon^{(d)}_{22}
  \end{array}
  \right) ,
  \end{aligned}
\end{equation}
with  
\begin{equation}
\varepsilon^{(u)}_{11} - 2\varepsilon^{(u)}_{12} 
+ \varepsilon^{(u)}_{22} = O(\lambda^4), \qquad
\varepsilon^{(d)}_{11} - 2\varepsilon^{(d)}_{12} 
+ \varepsilon^{(d)}_{22} = O(\lambda^2) .
\end{equation}
At any rate, we have a sufficient number of parameters to realize
the above fine-tuning.

\section{SUSY breaking terms}

Here we study SUSY breaking terms in the model with the $D_4$ flavor
structure. In particular, we are interested in the forms of sfermion
mass-squared matrices and $A$-terms, and study their degeneracies.

\subsection{Sfermion masses \label{sfermions}}

Let us first study squark masses. It is obvious that
before the $D_4$ flavor symmetry breaks the $D_4$ flavor structure
leads to the soft scalar mass-squared matrices,
\begin{equation}
m^2_{\phi_i \phi_j} =
\left(
\begin{array}{ccc}
m^2_{\phi_1 \phi_1} & 0 & 0 \\
0 & m^2_{\phi_1 \phi_1} &0 \\
0 & 0 & m^2_{\phi_3 \phi_3}
\end{array}
\right) ,
\label{scalar-mass}
\end{equation}
for $\phi_i = q_i, \ \bar u_i, \ \bar d_i$ ($i=1,2,3$).
However, we are interested in corrections to the above form from
$D_4$ flavor symmetry breaking, and we estimate such corrections
in what follows.

We consider the SUSY breaking scenario, where
moduli fields $M$ including the dilaton are dominant
in SUSY breaking \cite{BIM,multiT,BIMS}.
Such a scenario would be plausible within the framework of
string-inspired supergravity.

Before flavor symmetry breaking,
the $D_4$ flavor symmetry requires that
the K\"ahler potential of matter fields has the diagonal form,
\begin{equation}
K_{\rm matter} = \sum_{\phi_i=q_i, u_i, d_i}K_{\phi_i \phi^\dag_i}
(M)|\phi^i|^2, \end{equation} with
\begin{equation}
K_{\phi_1 \phi^\dag_1}(M) = K_{\phi_2 \phi^\dag_2}(M),
\end{equation}
where $\phi_i=q_i, u_i, d_i$.  In general, the K\"ahler metric
$K_{\phi_i \phi^\dag_i}$ depends on moduli fields $M$.

In general, we obtain the following soft SUSY breaking scalar masses
\cite{ST-soft},
\begin{equation}
m^2_{\phi_i \phi_i} = V_0 +m^2_{3/2} -
\sum_{a,b}F^{\Phi_a}\bar F^{\Phi_b}
\partial_{\Phi_a} \partial_{\bar \Phi_b} \ln (K_{\phi_i \phi^\dag_i}
),
\label{soft-mass}
\end{equation}
where $V_0$ is the vacuum energy
and $m_{3/2}$ is the gravitino mass defined by
the total K\"ahler potential $K$ and superpotential $W$
as $m^2_{3/2} \equiv \langle e^K|W|^2 \rangle$.
Thus, the $D_4$-
flavor structure leads to the soft scalar mass-squared matrices
(\ref{scalar-mass})
for $\phi_i = q_i, \ \bar u_i, \ \bar d_i$ ($i=1,2,3$).
Naturally, non-vanishing entries, $m^2_{\phi_i \phi_i}$ are
of $O(m^2_{3/2})$.

The $D_4$ breaking induces off-diagonal entries of the K\"ahler
metric and squark mass squared matrices. The (1,2) entry of the
K\"ahler metric for $\phi_i = q_i, \ \bar u_i, \ \bar d_i$ can be
induced by e.g.
\begin{equation}
C_{\phi_1 \phi_2^\dagger}(M)\sigma^{(d)}_{[D1]}\sigma^{(d) \dagger}_{[D2]}
\phi_1 \phi^\dagger_2 ,
\end{equation}
and other similar operators, where the coefficient
$C_{\phi_1 \phi_2^\dagger}(M)$ may depend on moduli $M$.
Similarly, the (2,1) entry can be induced.
Furthermore, the ($i$,3) and (3,$i$) entries ($i=1,2$)
for left-handed squarks can be induced by
\begin{equation}
C_{\phi_i \phi_3^\dagger}(M)\sigma^{(q)}_{[Di]}
q_i q^\dagger_3 ,\qquad
C_{\phi_3 \phi_i^\dagger}(M)\sigma^{(q) \dagger}_{[Di]}
q_3 q^\dagger_i.
\end{equation}
These corrections are dominant.
Although other terms are allowed, those are not
important in the following discussion.

This K\"ahler metric generates the following form of left-handed
squark masses squared in the flavor basis,
\begin{equation}
    m^2_q = \left(
      \begin{array}{ccc}
      m^2_{q_1 q_1} & O(\lambda^4m^2) &  O(\lambda^3m^2)\\
      O(\lambda^4m^2) & m^2_{q_1 q_1} & O(\lambda^2m^2) \\
      O(\lambda^3m^2) & O(\lambda^2m^2) & m^2_{q_3 q_3}
      \end{array}
      \right) ,
\end{equation}
where $m$ would be of the same order as $m_{q_1 q_1}$ and
$m_{q_3 q_3}$.
Similarly, down sector right-handed squark masses are obtained as
\begin{equation}
    m^2_d = \left(
      \begin{array}{ccc}
      m^2_{d_1 d_1} & O(\lambda^4m^2) &   0 \\
      O(\lambda^4m^2) & m^2_{d_1 d_1} &  0 \\
      0 & 0 & m^2_{d_3 d_3}
      \end{array}
      \right) ,
\end{equation}
where ($i$,3) and (3,$i$) entries are suppressed sufficiently.
The up sector right-handed squark masses have the same form.

Note that $F$-components $F_\sigma$ of $\sigma$ fields as well as
moduli $F$-terms contribute to squark masses. Both contributions
lead to the above form of squark masses squared, because we have
\begin{equation}
F_{\sigma[R]}^\dagger = -e^{ \langle K \rangle /2}
\langle K_{\sigma[R] \sigma[R]}
 \sigma_{[R]}^\dagger
 \hat W
+  \partial_{\sigma[R]} \hat W \rangle ,
\end{equation}
where $K_{\sigma[R] \sigma[R]}$ denotes the K\"ahler metric of
$\sigma_{[R]}$, and $\hat W$ is the non-trivial superpotential
leading to SUSY breaking, and naturally we estimate $F_\sigma /
\sigma =O(m_{3/2})$. $F$-components of $\sigma$-fields are more
important for estimating $A$-terms, and we will discuss them in more
detail in the next subsection.

Here we define mass insertion parameters $(\delta^{u,d}_{XY})_{ij}$, i.e.,
\begin{equation}
(\delta^{u,d}_{ij})_{XY} \equiv \frac{(m^{u,d}_{ij})^2_{XY}}{\tilde m^2},
\end{equation}
where $XY= LL, RR, LR$ and $\tilde m^2$ denotes the average squark
mass-squared.

Our model leads to
\begin{equation}
  \begin{aligned}
    \de{12}{LL} &\sim \lambda^4, &
    \de{13}{LL} &\sim \lambda^2, &
    \de{23}{LL} &\sim \lambda^2, \\
    \de{12}{RR} &\sim \lambda^4, &
    \de{13}{RR} &\lesssim \lambda^4, &
    \de{23}{RR} &\lesssim \lambda^4, \\
    \deu{12}{LL} &\sim \lambda^4, &
    \deu{13}{LL} &\sim \lambda^2, &
    \deu{23}{LL} &\sim \lambda^2, \\
    \deu{12}{RR} &\sim \lambda^4, &
    \deu{13}{RR} &\lesssim \lambda^4, &
    \deu{23}{RR} &\lesssim \lambda^4,
  \end{aligned}
\label{mass-in-LL-RR}
\end{equation}
at the Planck scale. In addition, we have flavor-blind
renormalization group (RG) effects due to gaugino masses of
$O(7M_{1/2}^2)$. Such RG effects reduce the above mass insertion
parameters by $O(10^{-1})$, because gaugino masses $M_{1/2}$ are
naturally of $O(m_{3/2})$ within the framework of dilaton/moduli
mediation. These values of mass insertion parameters satisfy
experimental constraints on FCNCs \cite{Gabbiani:1996hi,fcnc}.
When we derived Eq.~(\ref{mass-in-LL-RR}), we assumed that 
$m_{11}^2 / m_{33}^2 \sim O(1)$.
If the ratio is larger than $O(1/\lambda)$, then the mass insertion
parameters get enhanced by $1/\lambda$. Still they are phenomenologically
acceptable.

Furthermore, we have assumed extra $U(1)$ symmetries to obtain realistic
Yukawa matrices. Breaking of such extra symmetries, in general,
induces $D$-term contributions to soft scalar masses which are proportional
to the $U(1)$ charges of fields, as shown in Eqn.(\ref{D-term-mass}) of
Appendix \ref{app:U1FN}. (See Ref.~\cite{KK} for heterotic models.)
However, as a consequence of the $D_4$ flavor structure, such $D$-term
contributions are also degenerate between the first and second families,
because they must have the same $U(1)$ charges. Thus, after including
such D-term contributions, SUSY breaking scalar mass-squared matrices are
still of the form of Eqn.(\ref{scalar-mass}).
\footnote{Moreover, RG effects due to extra $U(1)$ gaugino masses are
also significant \cite{Kobayashi:2002mx}.  Such RG effects are also
degenerate between the first and second families in our model, because
they have the same $U(1)$ charges.}

\subsection{A-terms \label{aterms}}

Now, let us study the SUSY breaking
trilinear scalar couplings, i.e. the A-terms. 
Soft trilinear terms are obtained as \cite{ST-soft}
\begin{eqnarray}
h_{ijk} &=& \sum_{m}F^m [ \partial_m Y_{ijk}+\partial_m\hat K - 
 \sum_{n,p} ( 
K^{\phi_n \phi_p^\dagger} \partial_m K_{\phi_i \phi_p^\dagger} Y_{njk}
\nonumber \\
& & 
+ K^{\phi_n \phi_p^\dagger} \partial_m K_{\phi_j \phi_p^\dagger} Y_{ink}
+ K^{\phi_n \phi_p^\dagger} \partial_m K_{\phi_k \phi_p^\dagger}Y_{ijn}
) ],
\label{generic-A}
\end{eqnarray}
in generic case with non-vanishing off-diagonal elements 
of the K\"ahler metric.
When the K\"ahler metric of matter fields
(within the framework of supergravity) is diagonal, the SUSY breaking
trilinear scalar couplings are written as 
\begin{eqnarray}
h_{ijk} &=& Y_{ijk} A_{ijk} , \label{diagonal-A-1}\\
A_{ijk} &=& A^{(K)}_{ijk} + A^{(Y)}_{ijk},
\label{diagonal-A-2}
\end{eqnarray}
where
\begin{eqnarray}
A^{(K)}_{ijk} &=& \sum_m F^m [\partial_m \hat K -\partial_m
\ln(K_{\phi_i \phi_i^\dagger} K_{\phi_j \phi_j^\dagger} K_{\phi_k \phi_k^\dagger}) ],
\label{AK} \\
A^{(Y)}_{ijk} &=& \sum_m F^m \partial_m \ln(Y_{ijk}). \label{AY}
\end{eqnarray}
The first term of $A^{(K)}_{ijk}$ is the universal contribution due to
the K\"ahler potential $\hat K$ of the dilaton and moduli fields.
The second term of $A^{(K)}_{ijk}$ is the contribution through the wave
function, i.e. the K\"ahler metric.  The term $A^{(Y)}_{ijk}$ appears only
when Yukawa couplings are field-dependent. Prior to $D_4$ symmetry breaking,
the only non-vanishing entry in the Yukawa matrix is for $Y_{ijk}$ with
$i=j=3$ and $k = H$ (Higgs).

We consider the A-matrices after the $D_4$ symmetry breaking.
As studied in Section \ref{sfermions}, the $D_4$ symmetry breaking
induces off-diagonal elements of K\"ahler metric, e.g. for
left-handed squarks,
\begin{equation}
\left(
\begin{array}{ccc}
O(1) & O(\lambda^4) & O(\lambda^3) \\
O(\lambda^4) & O(1) & O(\lambda^2) \\
O(\lambda^3) & O(\lambda^2) & O(1)
\end{array}
\right) .   \label{eq:kahlermetric}
\end{equation}
Right-handed squarks of up and down sectors have a similar form of
K\"ahler metric, but ($i$,3) and (3,$i$) elements for $i=1,2$ are
more suppressed. The K\"ahler metric is almost diagonal, and such
diagonal form is violated by $O(\lambda^4)$ in the (1,2) entry.
Thus, it is reasonable to neglect off-diagonal elements of the
K\"ahler metric in the first approximation.  Then, as
we discuss shortly, our model of Yukawa matrices discussed in the
previous section leads to the following form of scalar trilinear
coupling matrices,
\begin{equation}
  \begin{aligned}
    h^{(u)} &= \left(
      \begin{array}{ccc}
        O( \lambda^4) & O( \lambda^4) & O(\lambda^3) \\
        O(\lambda^4) & O(\lambda^4) & O(\lambda^2) \\
        0       & 0       & O(1)
      \end{array}
    \right) \times A, \\
    h^{(d)} &= \left(
      \begin{array}{ccc}
        O(\lambda^3) & O(\lambda^3) & O(\lambda^3) \\
        O(\lambda^3) & O(\lambda^3) & O(\lambda^2) \\
        0       & 0       & O(1)
      \end{array}
    \right) \times \lambda \times A .
  \end{aligned}   \label{eq:trilinear}
\end{equation}
This seems to lead to mass insertion parameters, e.g.
\begin{equation}
\de{12}{LR} \sim \lambda^3 \times m_b A/\widetilde{m}^2.
\label{eq:deltaLR}
\end{equation}
However, due to quark-squark mass alignment, we will
show that our model actually leads to much smaller values of mass
insertion parameters.

In the following we derive the scalar tri-linear
couplings in Eq.~(\ref{eq:trilinear}). For the moment, let us
neglect off-diagonal elements of the K\"ahler metric as discussed
above. Since the $D_4$ flavor structure requires that $K_{\phi_1
\phi_1^\dagger}=K_{\phi_2 \phi_2^\dagger}$, the form of $A^{(K)}$
including the Higgs field is obtained as
\begin{equation}
A^{(K)}_{ijH} = \left(
\begin{array}{ccc}
A^{(K)}_0 &  A^{(K)}_0 &  A^{(K)}_1 \\
A^{(K)}_0 &  A^{(K)}_0 &  A^{(K)}_1 \\
A'^{(K)}_1 & A'^{(K)}_1 & A^{(K)}_3
\end{array}
\right) ,
\label{A-1}
\end{equation}
for both the up and down sectors.

Now, let us discuss the $A^{(Y)}$ part. The Yukawa couplings depend
on gauge-singlets $\sigma$ (\ref{yukawa-2}).  Thus, their
F-components $F^\sigma$ contribute to $A^{(Y)}$, and it is important
to evaluate $F^\sigma$. Within the framework of supergravity, the
F-component of $\sigma$ is written as
\begin{equation}
F_{\sigma[R]}^\dagger = -e^{ \langle K \rangle /2} \langle
K_{\sigma[R] \sigma[R]} \sigma_{[R]}^\dagger \hat W +
\partial_{\sigma[R]} \hat W \rangle ,  \label{eq:fsigma}
\end{equation}
where $K_{\sigma[R] \sigma[R]}$ denotes the K\"ahler metric of
$\sigma_{[R]}$, and $\hat W$ is the non-trivial
superpotential leading to SUSY breaking.
The $D_4$ symmetry requires
\begin{equation}
K_{\sigma[D_1] \sigma[D_1]} = K_{\sigma[D_2] \sigma[D_2]},
\end{equation}
as well as $K_{\sigma[D_1] \sigma[D_2]} = K_{\sigma[D_2]
\sigma[D_1]}=0$.  Here we assume
that the non-trivial superpotential does not include $\sigma$. In
this case, we can write 
\begin{equation}
\frac{F^{\sigma[R]}}{\langle \sigma_{[R]} \rangle}
 = -e^{ \langle K \rangle /2} \langle  \hat W^* \rangle.
\end{equation}
Therefore, the total A-matrices of the up and down
sectors have the form
\begin{equation}
A^{(u,d)}_{ijH} = \left(
\begin{array}{ccc}
A^{(u,d)}_0 &  A^{(u,d)}_0 &  A^{(u,d)}_1 \\
A^{(u,d)}_0 &  A^{(u,d)}_0 &  A^{(u,d)}_1 \\
A'^{(u,d)}_1 & A'^{(u,d)}_1 & A^{(u,d)}_3
\end{array}
\right) .
\label{A-2}
\end{equation}
The $(2 \times 2)$ sub-matrices for the first and second families
are degenerate. That implies that when we write Yukawa matrices of
our model as,
\begin{equation}
  \begin{aligned}
      y^{(u)} &= \left(
  \begin{array}{ccc}
    \cu{11} \lambda^4 & \cu{12} \lambda^4 & \cu{13} \lambda^3 \\
    \cu{21} \lambda^4 & \cu{22} \lambda^4 & \cu{23} \lambda^2 \\
    0       & 0       & 1
  \end{array}
  \right) , \\
      y^{(d)} &= \left(
  \begin{array}{ccc}
    \cd{11} \lambda^3 & \cd{12} \lambda^3 & \cd{13} \lambda^3 \\
    \cd{21} \lambda^3 & \cd{22} \lambda^3 & \cd{23} \lambda^2 \\
    0       & 0       & 1
  \end{array}
  \right) \times \lambda ,
  \end{aligned}
\end{equation}
in the $D_4$ basis, the scalar trilinear coupling matrices have the
following form,
\begin{equation}
  \begin{aligned}
    h^{(u)} &= \left(
      \begin{array}{ccc}
        \cu{11} \lambda^4 & \cu{12} \lambda^4 & b^{(u)}\,\cu{13} \lambda^3 \\
        \cu{21} \lambda^4 & \cu{22} \lambda^4 & b^{(u)}\,\cu{23} \lambda^2 \\
        0       & 0       & c^{(u)}
      \end{array}
    \right) \times A, \\
    h^{(d)} &= \left(
      \begin{array}{ccc}
        \cd{11} \lambda^3 & \cd{12} \lambda^3 & b^{(d)}\,\cd{13} \lambda^3 \\
        \cd{21} \lambda^3 & \cd{22} \lambda^3 & b^{(d)}\,\cd{23} \lambda^2 \\
        0       & 0       & c^{(d)}
      \end{array}
    \right) \times \lambda \times A .
  \end{aligned}
\end{equation}
as given in Eq.~(\ref{eq:trilinear}).  Note, this form
is quite different from one which is obtained in the U(1)
Froggatt-Nielsen model as shown in Appendix B.

Now consider the consequence of quark-squark mass
alignment. The upper left $2\times2$ sub-matrices of $y^{(u)}$
($y^{(d)}$) and $h^{(u)}$ ($h^{(d)}$) are proportional to each
other. We can multiply each of $y^{(u)}$ and $y^{(d)}$ by two
unitary matrices on the left and the right hand sides to diagonalize
the upper left $2\times2$ sub-matrix. After doing this, we obtain
\begin{equation}
  \begin{aligned}
      y^{(u)} &= \left(
  \begin{array}{ccc}
    \cpu{11} \lambda^7 & 0                 & \cpu{13} \lambda^2 \\
    0                 & \cpu{22} \lambda^4 & \cpu{23} \lambda^2 \\
    0       & 0       & 1
  \end{array}
  \right) , \\
      y^{(d)} &= \left(
  \begin{array}{ccc}
    \cpd{11} \lambda^4 & 0                 & \cpd{13} \lambda^2 \\
    0                 & \cpd{22} \lambda^3 & \cpd{23} \lambda^2 \\
    0       & 0       & 1
  \end{array}
  \right) \times \lambda .
  \end{aligned}
\end{equation}
In the same basis, the $h$ matrices look like
\begin{equation}
  \begin{aligned}
    h^{(u)} &= \left(
      \begin{array}{ccc}
        \cpu{11} \lambda^7 & 0                 & b^{(u)}\,\cpu{13} \lambda^2 \\
        0                 & \cpu{22} \lambda^4 & b^{(u)}\,\cpu{23} \lambda^2 \\
        0       & 0       & c^{(u)}
      \end{array}
    \right) \times A, \\
    h^{(d)} &= \left(
      \begin{array}{ccc}
        \cpd{11} \lambda^4 & 0                 & b^{(d)}\,\cpd{13} \lambda^2 \\
        0                 & \cpd{22} \lambda^3 & b^{(d)}\,\cpd{23} \lambda^2 \\
        0       & 0       & c^{(d)}
      \end{array}
    \right) \times \lambda \times A .
  \end{aligned}
\end{equation}
Then, we can estimate the $LR$ and $RL$ mass insertion parameters
applying the perturbative diagonalization formula to the above matrices,
\begin{equation}
  \begin{aligned}
    \de{12}{LR} &\sim \lambda^7 \times m_b A/\widetilde{m}^2, &
    \de{13}{LR} &\sim \lambda^2 \times m_b A/\widetilde{m}^2, &
    \de{23}{LR} &\sim \lambda^2 \times m_b A/\widetilde{m}^2, \\
    \de{12}{RL} &\sim \lambda^8 \times m_b A/\widetilde{m}^2, &
    \de{13}{RL} &\sim \lambda^6 \times m_b A/\widetilde{m}^2, &
    \de{23}{RL} &\sim \lambda^5 \times m_b A/\widetilde{m}^2, \\
    \deu{12}{LR} &\sim \lambda^8 \times m_t A/\widetilde{m}^2, &
    \deu{13}{LR} &\sim \lambda^2 \times m_t A/\widetilde{m}^2, &
    \deu{23}{LR} &\sim \lambda^2 \times m_t A/\widetilde{m}^2, \\
    \deu{12}{RL} &\sim \lambda^{11} \times m_t A/\widetilde{m}^2, &
    \deu{13}{RL} &\sim \lambda^9 \times m_t A/\widetilde{m}^2, &
    \deu{23}{RL} &\sim \lambda^6 \times m_t A/\widetilde{m}^2.
  \end{aligned}
\label{mass-in-LR}
\end{equation}
As noted previously, these are much smaller than their
naive value, Eq.~(\ref{eq:deltaLR}). They satisfy  experimental
constraints \cite{Gabbiani:1996hi,fcnc}. When we derived 
Eq.~(\ref{mass-in-LR}),
we assumed that all the $A$'s in Eq.~(\ref{A-2}) have the same sizes, and
all of their ratios are of $O(1)$. If their ratios are larger than
$O(1/\lambda)$, then the above estimate should be multiplied by
$O(1/\lambda)$, which is still phenomenologically viable.

Up until now we have neglected off-diagonal elements of the K\"ahler
metric. Here we discuss corrections due to these neglected terms.
Such corrections violate the degeneracy of the upper left $(2 \times
2)$ sub-matrices in $A^{(u,d)}_{ijH}$ (\ref{A-2}) by $O(\lambda^4)$.
Similarly, they can make corrections to other entries.
Note that Eqs.~(\ref{diagonal-A-1})-(\ref{AY}) are not available 
for non-vanishing off-diagonal elements of the K\"ahler
metric and we have to use the generic formula (\ref{generic-A}).
Including these corrections modifies the scalar trilinear couplings to
\begin{equation}
  \begin{aligned}
    h^{(u)} &= \left(
      \begin{array}{ccc}
        [\cu{11}+O(\lambda^4)]\lambda^4&[\cu{12}+O(\lambda^4)]\lambda^4&
        [b^{(u)}\,\cu{13}+O(\lambda^3)]\lambda^3 \\
      {}[\cu{21}+O(\lambda^4)]\lambda^4&[\cu{22}+O(\lambda^4)]\lambda^4&
        [b^{(u)}\,\cu{23}+O(\lambda^4)]\lambda^2 \\
       O(\lambda^6)  & O(\lambda^6) & c^{(u)}
      \end{array}
    \right) \times A, \\
    h^{(d)} &= \left(
      \begin{array}{ccc}
        [\cd{11}+O(\lambda^4)]\lambda^3&[\cd{12}+O(\lambda^4)]\lambda^3&
        [b^{(d)}\,\cd{13}+O(\lambda^3)]\lambda^3 \\
      {}[\cd{21}+O(\lambda^4)]\lambda^3&[\cd{22}+O(\lambda^4)]\lambda^3&
        [b^{(d)}\,\cd{23}+O(\lambda^4)] \lambda^2 \\
        O(\lambda^6)       & O(\lambda^6)       & c^{(d)}
      \end{array}
    \right) \times \lambda \times A .
  \end{aligned}
\end{equation}
Corrections such as these do not drastically change the above
estimation of mass insertion parameters.

\section{Conclusions}

In this paper we have discussed the structure of heterotic string
models with the discrete non-abelian flavor symmetry $D_4$. Such
flavor symmetries are easily obtained in a variety of orbifold
constructions of the heterotic string
\cite{Kobayashi:2004ud,Kobayashi:2004ya,Forste:2004ie,Kobayashi:2006wq}.
For example, $D_4$ flavor symmetries are easily obtained in $Z_2
\times Z_N$ orbifold constructions.   We have also shown that these
discrete non-abelian flavor symmetries may be useful for
understanding the hierarchy of fermion masses and mixing angles. In
addition, they constrain the SUSY breaking mass-squared matrices and
cubic scalar interaction matrices; hence suppressing flavor
violating processes. In particular, the non-abelian flavor
symmetries, with quarks of the first two families in one irreducible
representation, reduce the sensitivity to flavor violating
interactions.  In Appendix \ref{app:U1FN}, we have also compared the
structure of soft SUSY breaking mass-squared matrices and A-term
matrices consistent with discrete non-abelian flavor symmetries and
those of U(1) flavor symmetries.   As discussed, U(1) flavor
symmetries do not sufficiently constrain FCNC processes.

We have considered only the quark sector.  We can extend the
previous analysis to the lepton sector. Indeed, we can obtain the
same results for Yukawa matrices, SUSY breaking scalar masses and
A-terms as eqs.(\ref{yukawa-2}),(\ref{scalar-mass}),(\ref{A-2}).
With a $D_4$ flavor symmetry, there is no difficulty encountered for
obtaining the standard See-Saw mechanism with heavy right-handed
neutrinos.  However there are still many more additional parameters
due to the right-handed Majorana masses.  The number of free
parameters is typically larger than the number of observables. Thus,
we have no prediction in the Yukawa sector, unless we impose texture
zeros or assume additional symmetries.  On the other hand, we have
predictions for SUSY breaking terms. Actually, the branching ratio
for $\mu \rightarrow e \gamma$ leads to the strongest constraint.
Hence, the degeneracy between the first and second families would
help to satisfy this constraint, in particular by suppressing the
factor, $(\delta^\ell_{LL})_{12}$.

In order to obtain more predictive Yukawa sectors we need to reduce
the number of free parameters,  This can be done by embedding the
flavor structure into GUTs.  In some orbifold GUT and string models
with an intermediate Pati-Salam gauge symmetry, the third generation
Yukawa couplings are unified with $\lambda_t = \lambda_b =
\lambda_\tau = \lambda_{\nu_\tau}$. However in order for these
theories to be phenomenologically acceptable, certain relations
among the soft SUSY breaking terms must hold,  in particular $A_{33}
\approx - 2 m_{33}$ \cite{bdr}.  It is clear that this relation is
not satisfied with simple scenarios of dilaton or moduli SUSY
breaking. A possible explanation may require a more complicated SUSY
breaking scenario, for example see Appendix \ref{app:so10}.

\acknowledgements

We would like to acknowledge R.-J. Zhang who participated during the
early stages of this work. T.~K.\/ is supported in part by the
Grand-in-Aid for Scientific Research \#17540251 and the Grant-in-Aid
for the 21st Century COE ``The Center for Diversity and Universality
in Physics'' from the Ministry of Education, Culture, Sports,
Science and Technology of Japan. S.R. acknowledges partial support
under DOE contract DOE/ER/01545-865.
JhP was supported by
the JSPS postdoctoral fellowship program
for foreign researchers and the accompanying grand-in-aid no.\ 17.05302.
PK was supported by KOSEF SRC program through CHEP at Kyungpook National
University.

\appendix

\section{$D_4$ discrete group}

The $D_4$ discrete group has five representations including a doublet $D$,
a trivial singlet $A_1$ and three
non-trivial singlets $B_1,B_2,A_2$, which are shown in Table 2.

\begin{table}[htb]
\begin{center}
\begin{tabular}{|c|c|c|c|c|c|}  \hline
Representations & $I$ & $-I$ & $\pm \sigma_1$ &
$\pm \sigma_3$ & $\mp i\sigma_2$ \\ \hline \hline
Doublet$-D$ & 2 & --2 & 0 & 0 & 0 \\ \hline
Singlet$-A_1$ & 1 & 1 & 1 & 1 & 1 \\ \hline
Singlet$-B_1$ & 1 & 1 & 1 & --1 & --1 \\ \hline
Singlet$-B_2$ & 1 & 1 & --1 & 1 & --1 \\ \hline
Singlet$-A_2$ & 1 & 1 & --1 & --1 & 1 \\ \hline
\end{tabular}
\end{center}
\caption{Representations of $D_4$ symmetry}
\end{table}

A product of two doublets is decomposed as four singlets,
\begin{equation}
(D \times D) = A_1 + B_1 + B_2 + A_2 .
\end{equation}
More explicitly, we consider two
$D_4$ doublets $S_A$ and $\bar S_A$ $(A=1,2)$.
Their product $S_A \bar S_B$ is decomposed
in terms of  $A_1,B_1,B_2,A_2$,
\begin{eqnarray}
S_1 \bar S_1 + S_2 \bar S_2 &\sim & A_1 ,\\
S_1 \bar S_2 + S_2 \bar S_1 &\sim & B_1 ,\\
S_1 \bar S_1 - S_2 \bar S_2 &\sim & B_2 ,\\
S_1 \bar S_2 - S_2 \bar S_1 &\sim & A_2 .
\end{eqnarray}

\section{SUSY breaking terms in the U(1) Froggatt-Nielsen model \label{app:U1FN}}

One of famous flavor mechanisms is the U(1) Froggatt-Nielsen mechanism
in the string-inspired approach. Here we give a brief comments on soft
SUSY breaking terms to compare them with our results from the $D_4$ flavor
structure.

In the simple U(1) FN mechanism, the effective Yukawa couplings are
obtained through the higher dimensional operators, e.g.
\begin{equation}
\chi^{q_{Qi}+q_{uj}+q_{Hu,d}} Q_i u_j
H_{u,d},
\end{equation}
where $\chi$ is the FN field with non-vanishing VEV and $q_{Qi}, q_{uj}$
and $ q_{Hu,d}$ are extra U(1) charges of $Q_i, u_j$ and $H_{u,d}$,
respectively. In this mechanism, it is easy to derive realistic Yukawa
matrices.

We have the usual sfermion masses due to F-components of moduli fields
like Eqs.~(\ref{soft-mass}).
Since the U(1) flavor symmetry, in general, does not specify the
K\"ahler metric of matter fields, one can not give a generic statement
on this part. In addition to this usual part, the extra U(1) breaking
induces additional contributions to sfermion masses, that is, the so-called
$D$-term contributions. Such  part is, in general, written as
\begin{equation}
\Delta m^2_\alpha = q_\alpha m_D^2,
\label{D-term-mass}
\end{equation}
where $q_\alpha$ is the extra U(1) charge of matter field, and $m^2_D$ itself
is the universal for all matter fields, that is, these are proportional
to extra U(1) charges. (See Ref.~\cite{KK} for heterotic models.)
Furthermore, RG effects due to extra $U(1)$ gaugino masses may
generate significant non-degeneracy when each family has different
$U(1)$ charges \cite{Kobayashi:2002mx}.

As Section \ref{aterms}, A-terms are obtained by calculating eqs.
~(\ref{AK}), (\ref{AY}). In this case, we obtain
the $A^{(Y)}$-matrices \cite{Abel:2001cv,Ross:2002mr},
\begin{equation}
A_{ijH} = \frac{F^\chi}{\chi}(q_{Qi}+q_{uj}+q_H),  \label{eq:A}
\end{equation}
for $Y_{ij H} =  \chi^{q_{Qi}+q_{uj}+q_H} $.  Here note that
$\frac{F^\chi}{\chi} =O(m_{3/2})$ because $F^\chi = \chi \bar W$.
Also note that the flavor-dependence in the second term of $A^{(K)}$
can be separated.
As a result, A-terms are decomposed as \cite{Kobayashi:2000br}
\begin{equation}
A_{\alpha \beta H} = A^{L}_\alpha + A^{R}_\beta.
\end{equation}
Moreover, the trilinear scalar coupling matrix $h_{i j H}$
is written as
\begin{equation}
h_{ij H} = Y_{ij } A_{ij H}
=\left(
\begin{array}{ccc}
 & & \\
 & Y_{ij } & \\
 & & \\
\end{array}
\right)
\cdot
\left(
\begin{array}{ccc}
A^R_1 & & \\
 & A^R_2 & \\
 & & A^R_3 \\
\end{array}
\right)
+\left(
\begin{array}{ccc}
A^L_1 & & \\
 & A^L_2 & \\
 & & A^L_3 \\
\end{array}
\right)
\cdot
\left(
\begin{array}{ccc}
 & & \\
 & Y_{ij } & \\
 & & \\
\end{array}
\right).
\end{equation}
This form of the A-matrices, in general, leads to dangerous FCNC
effects.

\section{SO(10) Yukawa unification and SUSY breaking terms \label{app:so10}}

SO(10) Yukawa unification for the third family, resulting from the
renormalizable coupling
\begin{equation}
W \supset \lambda \ 16_3 \ 10 \ 16_3 ,
\end{equation}
gives
\begin{equation}
\lambda_t = \lambda_b = \lambda_\tau = \lambda_{\nu_\tau} = \lambda .
\end{equation}
In order to fit the top, bottom and tau masses at the weak scale,
it has been shown that it is necessary to be in a particular region of
soft SUSY breaking parameter space~\cite{bdr}.
Define the soft SUSY breaking parameters:
$A_0$, the cubic scalar interaction mass term;
$M_{1/2}$, a universal gaugino mass;  $m_{16}$, the soft scalar
mass for squarks and sleptons; and $m_{10}$,
the Higgs soft SUSY breaking mass.   Then we require the relation
\begin{eqnarray}
A_0 & = & - 2 \ m_{16} \label{eq:A0} \\
                  m_{10} & = & \sqrt{2} \ m_{16}
\end{eqnarray}
and
\begin{equation}
\mu \sim M_{1/2} \ll m_{16} .
\end{equation}

The question is can this relation come naturally in string theory.
It is difficult to obtain this result from a combination of dilaton
and T moduli SUSY breaking.  Here we argue that this simple relation can
come from D-term and MSSM singlet SUSY breaking.  Consider the $U(1)_X$
symmetry in $E_6$ which commutes with $SO(10)$. The {\bf 27} dimensional
representation of $E_6$ decomposes under $SO(10) \times U(1)_X$ as
\begin{equation}
{\bf 27} \rightarrow  ({\bf 16, 1}) \oplus ({\bf 10, -2}) \oplus ({\bf 1, 4}) .
\end{equation}
If we now assume the {\bf 16} of quarks and leptons comes from a {\bf 27},
while the Higgs doublets come from a $\overline{\bf 27}$, we obtain
the wanted relation
\begin{equation}
m_{10}^2 = 2 \ m_{16}^2 \equiv 2 \ D_X .
\end{equation}
We also obtain the same scalar mass for all three families of squarks
and sleptons, consistent with minimal flavor violation.  Furthermore,
the relation
\begin{equation}
\mu \sim M_{1/2} \ll  m_{16}
\end{equation}
is easy to accommodate, for example by subdominant dilaton SUSY breaking.
Thus the only remaining question is the origin of the cubic scalar parameter
$A_0$.

Assume we have a term in the superpotential of the form
\begin{equation}
W \supset \chi^{Q_X(Q_i)+Q_X(u_j)+Q_X(H_{u,d})} \ Q_i \ u_j \ H_{u,d} =
\chi^4 \ Q_i \ u_j \ H_{u,d},
\end{equation}
where $\chi$ (with $Q_X(\chi) = -1$) is an MSSM singlet field.
Then in order to obtain the relation (Eqn.\ref{eq:A0}) from
Eqn. \ref{eq:A} we need
\begin{equation}
A_0 = 4 F_\chi/\chi \approx  - 2 \ \sqrt{D_X} .
\end{equation}
This relation may also be  accommodated.

\end{document}